\documentclass{elsart}
\usepackage{graphicx,amsmath,amssymb}

\begin{document}

\def\la{\; \raise0.3ex\hbox{$<$\kern-0.75em\raise-1.1ex\hbox{$\sim$}}\;}
\def\ga{\;  \raise0.3ex\hbox{$>$\kern-0.75em\raise-1.1ex\hbox{$\sim$}}\;}
\def\pFn{p_{\raise-0.3ex\hbox{{\scriptsize F$\!$\raise-0.03ex\hbox{\rm n}}}}
}  

\def\pFa{p_{\raise-0.3ex\hbox{{\scriptsize F$\!$\raise-0.03ex\hbox{$\alpha$}}}}
}  

\def\pFas{p_{\raise-0.3ex\hbox{{\scriptsize F$\!$\raise-0.03ex\hbox{$\alpha
'$}}}}
}  

\def\pFb{p_{\raise-0.3ex\hbox{{\scriptsize F$\!$\raise-0.03ex\hbox{$\beta$}}}}
}  

\def\vFa{v_{\raise-0.3ex\hbox{{\scriptsize F$\!$\raise-0.03ex\hbox{$\alpha$}}}}
}  
\def\pFp{p_{\raise-0.3ex\hbox{{\scriptsize F$\!$\raise-0.03ex\hbox{\rm p}}}}
}  
\def\pFe{p_{\raise-0.3ex\hbox{{\scriptsize F$\!$\raise-0.03ex\hbox{\rm e}}}}
}  
\def\pFmu{p_{\raise-0.3ex\hbox{{\scriptsize F$\!$\raise-0.03ex\hbox{\rm
$\mu$}}}} }  
\def\m@th{\mathsurround=0pt }
\def\eqalign#1{\null\,\vcenter{\openup1\jot \m@th
   \ialign{\strut$\displaystyle{##}$&$\displaystyle{{}##}$\hfil
   \crcr#1\crcr}}\,}

\newcommand{\vp}{\mbox{${\pmb p}$}}         
\newcommand{\vps}{\mbox{${\vp '}$}}         
\newcommand{\vQ}{\mbox{${\pmb Q}$}}         
\newcommand{\vQa}{\mbox{$\vQ_{\alpha}$}}         

\newcommand{\ap}{\mbox{$a_{\pmb p}^{(\alpha)}$}}
\newcommand{\apc}{\mbox{$a_{\pmb p}^{(\alpha) \dagger}$}}
\newcommand{\apQ}{\mbox{$a_{{\pmb p}+{\pmb Q}_{\alpha}}^{(\alpha)}$}}
\newcommand{\apcQ}{\mbox{$a_{{\pmb p}+{\pmb Q}_{\alpha}}^{(\alpha) \dagger}$}}
\newcommand{\bp}{\mbox{$b_{\pmb p}^{(\alpha)}$}}
\newcommand{\bpc}{\mbox{$b_{\pmb p}^{(\alpha) \dagger}$}}
\newcommand{\bpQ}{\mbox{$b_{{\pmb p}+{\pmb Q}_{\alpha}}^{(\alpha)}$}}
\newcommand{\bpcQ}{\mbox{$b_{{\pmb p} +{\pmb Q}_{\alpha}}^{(\alpha) \dagger}$}}

\newcommand{\aps}{\mbox{$a_{{\pmb p} '}^{(\alpha')}$}}
\newcommand{\apsc}{\mbox{$a_{{\pmb p} '}^{(\alpha') \dagger}$}}
\newcommand{\apsQ}{\mbox{$a_{{\pmb p} '+{\pmb Q}_{\alpha'}}^{(\alpha')}$}}
\newcommand{\apscQ}{\mbox{$a_{{\pmb p}'+{\pmb Q}_{\alpha'}}^{(\alpha')
\dagger}$}}
\newcommand{\bps}{\mbox{$b_{{\pmb p}  '}^{(\alpha')}$}}
\newcommand{\bpsc}{\mbox{$b_{{\pmb p} '}^{(\alpha') \dagger}$}}
\newcommand{\bpsQ}{\mbox{$b_{{\pmb p}  '+{\pmb Q}_{\alpha'}}^{(\alpha')}$}}
\newcommand{\bpscQ}{\mbox{$b_{{\pmb p} ' +{\pmb Q}_{\alpha'}}^{(\alpha')
\dagger}$}}

\newcommand{\thp}{\mbox{$\theta_{\pmb p}^{(\alpha)}$}}
\newcommand{\thpQ}{\mbox{$\theta_{{\pmb p}+{\pmb Q}_{\alpha}}^{(\alpha)}$}}
\newcommand{\thps}{\mbox{$\theta_{{\pmb p} '}^{(\alpha')}$}}
\newcommand{\thpsQ}{\mbox{$\theta_{{\pmb p} ' +{\pmb 
Q}_{\alpha'}}^{(\alpha')}$}}
\newcommand{\uup}{\mbox{$u_{\pmb p}^{(\alpha)}$}}
\newcommand{\vvp}{\mbox{$v_{\pmb p}^{(\alpha)}$}}
\newcommand{\uuup}{\mbox{$U_{\pmb p}^{(\alpha)}$}}
\newcommand{\vvvp}{\mbox{$V_{\pmb p}^{(\alpha)}$}}
\newcommand{\uups}{\mbox{$u_{{\pmb p}  '}^{(\alpha')}$}}
\newcommand{\vvps}{\mbox{$v_{{\pmb p}  '}^{(\alpha')}$}}
\newcommand{\uuups}{\mbox{$U_{{\pmb p}  '}^{(\alpha')}$}}
\newcommand{\vvvps}{\mbox{$V_{{\pmb p}  '}^{(\alpha')}$}}
\newcommand{\fp}{\mbox{$\mathfrak{f}_{\pmb p}^{(\alpha)}$}}
\newcommand{\np}{\mbox{$n_{\pmb p}^{(\alpha)}$}}
\newcommand{\fps}{\mbox{$\mathfrak{f}_{{\pmb p}  '}^{(\alpha')}$}}
\newcommand{\nps}{\mbox{$n_{{\pmb p}  '}^{(\alpha')}$}}
\newcommand{\mN}{\mbox{$\mathcal N$}}
\newcommand{\mF}{\mbox{$\mathcal F$}}
\newcommand{\vk}{\mbox{\boldmath $k$}}         
\newcommand{\xixi}{\mbox{\boldmath $\xi$}}         
\newcommand{\vq}{\mbox{\boldmath $q$}}         
\newcommand{\vr}{\mbox{\boldmath $r$}}         
\newcommand{\om}{\mbox{$\omega$}}              
\newcommand{\Om}{\mbox{$\Omega$}}              
\newcommand{\Th}{\mbox{$\Theta$}}              
\newcommand{\ph}{\mbox{$\varphi$}}             
\newcommand{\del}{\mbox{$\delta$}}             
\newcommand{\Del}{\mbox{$\Delta$}}             
\newcommand{\lam}{\mbox{$\lambda$}}            
\newcommand{\Lam}{\mbox{$\Lambda$}}            
\newcommand{\ep}{\mbox{$\varepsilon$}}         
\newcommand{\ka}{\mbox{$\kappa$}}              
\newcommand{\dd}{\mbox{d}}                     
\newcommand{\vect}[1]{\bf #1}                
\newcommand{\vtr}[1]{\mbox{\boldmath $#1$}}  
\newcommand{\vF}{\mbox{$v_{\mbox{\raisebox{-0.3ex}{\scriptsize F}}}$}}  
\newcommand{\pF}{\mbox{$p_{\mbox{\raisebox{-0.3ex}{\scriptsize F}}}$}}  
\newcommand{\kF}{\mbox{$k_{\rm F}$}}           
\newcommand{\kTF}{\mbox{$k_{\rm TF}$}}         
\newcommand{\kB}{\mbox{$k_{\rm B}$}}           
\newcommand{\tn}{\mbox{$T_{{\rm c}n}$}}        
\newcommand{\tp}{\mbox{$T_{{\rm c}p}$}}        
\newcommand{\te}{\mbox{$T_{eff}$}}             
\newcommand{\ex}{\mbox{\rm e}}                 
\newcommand{\rate}{\mbox{${\rm erg~cm^{-3}~s^{-1}}$}}
\newcommand{\mur}{\raisebox{0.2ex}{\mbox{\scriptsize (í)}}} 
\newcommand{\Mn}{\raisebox{0.2ex}{\mbox{\scriptsize (í{\it n\/})}}}        %
\newcommand{\Mp}{\raisebox{0.2ex}{\mbox{\scriptsize (í{\it p\/})}}}        %
\newcommand{\MN}{\raisebox{0.2ex}{\mbox{\scriptsize (í{\it N\/})}}}        %


\begin{frontmatter}
\title{The entrainment matrix of a superfluid 
neutron-proton mixture at a finite temperature}
\author{M.E.\ Gusakov$^{(a, b)}$\thanksref{MEG}} and
\author{P. Haensel$^{(b)}$\thanksref{PH}}
\address{$^{(a)}$Ioffe Physical Technical Institute, \\
Politekhnicheskaya 26, 194021 St.-Petersburg, Russia \\
$^{(b)}$N. Copernicus Astronomical Center, 
Bartycka 18, 00-716 Warsaw, Poland}

\thanks[MEG]{E-mail: gusakov@astro.ioffe.ru}
\thanks[PH]{E-mail: haensel@camk.edu.pl}
\begin{abstract}
\noindent
The entrainment matrix
(also termed the Andreev-Bashkin matrix
or the mass-density matrix)
for a neutron-proton mixture is derived
at a finite temperature in a neutron star core.
The calculation is performed in the frame of the
Landau Fermi-liquid theory generalized
to account for superfluidity of nucleons.
It is shown, that the temperature dependence
of the entrainment matrix
is described by a universal
function independent
on an actual model
of nucleon-nucleon interaction employed.
The results are presented in the form
convenient for their practical use.
The entrainment matrix is important,
e.g., in kinetics of superfluid nucleon mixtures
or in studies of the dynamical evolution of neutron stars
(in particular, in
the studies of star pulsations and pulsar glitches).
\vspace{0.5cm}

\noindent{\it PACS:}
21.65.+f;          
71.10.Ay;          
97.60.Jd;          
26.60.+c \\        
\noindent{\it Keywords:} Neutron star matter;
Fermi liquid theory; Superfluidity; Nucleon superfluid densities
\end{abstract}
\end{frontmatter}

\newpage

\section{Introduction}
\label{introduction}
It is well known, that
the neutron star core becomes superfluid (superconducting) at a
certain stage of neutron star thermal evolution (see, e.g.,
Ref. \cite{LS01}).
It is generally agreed that
protons pair in the spin singlet
($^1S_0$) state, while neutrons
pair in the spin triplet ($^3P_2$) state
in the neutron star core.
A variety of different models of nucleon
pairing have been proposed in literature
(references to original papers can be found
in Yakovlev et al. \cite{Yetal99} 
and in Lombardo and Schulze \cite{LS01}).
These models predict very different
density profiles of neutron (n) or proton (p)
critical temperatures $T_{c {\rm n, p}}(\rho)$.
In addition, it is not absolutely clear whether
the actual projection of
angular momentum of neutron pair onto
quantization axis is $m_J=0$
(as one usually assumes)
or it can be $|m_J|=1$ or 2.
For example, Amundsen and {\O}stgaard \cite{AO85}
found that the energetically preferable
state of a neutron pair can be a superposition
of states with different $m_J$.

Despite many theoretical uncertainties
it is obvious that superfluidity
strongly affects the evolution
of neutron stars, for example, its cooling
(see, e.g., Ref. \cite{YP04}),
neutron star pulsations
(see, e.g., Refs. \cite{M91a} -- \cite{Petal04})
and is probably related to pulsar glitches
(see Refs. \cite{Aetal84,Aetal03}).
 
One of the key ingredients of hydrodynamics and
kinetics of superfluid mixtures is the {\it entrainment matrix}
$\rho_{\alpha \alpha'}$ (also termed
the Andreev-Bashkin matrix or the mass-density matrix).
Implying, for simplicity, that the only baryons
in a neutron star core are neutrons and protons,
the matrix $\rho_{\alpha \alpha'}$
can be defined as \cite{AB75}:
\begin{eqnarray}
{\pmb J}_{\rm n} &=& (\rho_{\rm n} - \rho_{\rm nn}-
\rho_{\rm np}) \, {\pmb V}_{\rm qp} +
\rho_{\rm nn} \, {\rm {\pmb V}}_{{\rm n} s} +
\rho_{\rm np} \, {\rm {\pmb V}}_{{\rm p} s} \,,
\label{Jn} \\
{\pmb J}_{\rm p} &=&
(\rho_{\rm p} - \rho_{\rm pp}-
\rho_{\rm pn}) \, {\pmb V}_{\rm qp} +
\rho_{\rm pp} \, {\rm {\pmb V}}_{{\rm p} s} +
\rho_{\rm pn} \, {\rm {\pmb V}}_{{\rm n} s} \,.
\label{Jp}
\end{eqnarray}
Here $\rho_\alpha = m_{\alpha} n_{\alpha}$; 
$n_{\alpha}$ and $m_{\alpha}$ are the number density 
and the mass of nucleon species $\alpha={\rm n}$ or ${\rm p}$; 
${\pmb J}_{\alpha}$ and ${\pmb V}_{\alpha s}$ are the mass current
density and the superfluid velocity; ${\pmb V}_{\rm qp}$ is
the normal velocity of thermal excitations 
(see, e.g., Refs. \cite{LP80,K89}).
We assume that ${\pmb V}_{\rm qp}$
is the same for nucleons of both species.
Eqs. (\ref{Jn}) and (\ref{Jp})
differ from a ``natural''
expression for the mass current density
${\pmb J}_{\alpha} = \rho_{\alpha} {\pmb V}_{\alpha}$
(with ${\pmb V}_{\alpha}$ being the momentum per unit mass
of nucleon species $\alpha$) for two reasons.

First, three independent motions can exist
in a mixture of two
superfluids, each carrying a mass
(see, e.g., Ref. \cite{K73}).
They are the motion of thermal excitations with
the velocity ${\pmb V}_{{\rm qp}}$ and two superfluid
motions with the velocities ${\pmb V}_{{\rm n} s}$ and ${\pmb V}_{{\rm p} s}$.

Second, the superfluid flow of one component of the mixture
entrains a flow of another component and vice versa.
For example, the superfluid motion of neutrons
carries along some part of the mass of protons
because nucleon liquids are strongly interacting
(see, e.g., Ref. \cite{S76}).
The non-diagonal elements $\rho_{\rm np}$ and
$\rho_{\rm pn}$, therefore, characterize
the intensity of neutron-proton coupling.
In particular, in the absence of interactions between
neutrons and protons one has $\rho_{\rm np}=\rho_{\rm pn}=0$.

It follows from the phenomenological
analysis of Andreev and Bashkin \cite{AB75}
that the entrainment matrix should be symmetric
($\rho_{\rm np}=\rho_{\rm pn}$),
since it can be presented in the form:
\begin{equation}
\rho_{\alpha \alpha'} = \left( \frac{\partial^2 E}{\partial {\pmb V}_{\alpha s}
\partial {\pmb V}_{\alpha' s}} \right)_{n_{\rm n}, n_{\rm p}, T},
\label{AB}
\end{equation}
where $E$ is the energy density
in the coordinate frame in which
${\pmb V}_{\rm qp}=0$, and $T$ is the temperature.

The entrainment matrix $\rho_{\alpha \alpha'}$
of a neutron-proton mixture 
was calculated by Borumand et al. \cite{Betal96}
and Comer and Joynt \cite{CJ03} for $T=0$.
However, in many cases the
zero-temperature approximation cannot be justified.
For example, one needs the
matrix $\rho_{\alpha \alpha'}$
at non-zero temperatures
for analyzing kinetic properties of matter,
especially the kinetic coefficients
(the bulk and shear viscosity, the diffusion coefficient).
Also, the entrainment matrix is required at $T \neq 0$
for investigating pulsations of warm neutron stars or
stars possessing the pulsation energy
of the order of or higher
than its thermal energy
(It is implied that pulsation energy,
being dissipated,
is able to heat the star substantially
and thus to change $\rho_{\alpha \alpha'}$.
A simple example, illustrating
the influence of temperature related effects
on the neutron star pulsations,
was considered by Gusakov et al. \cite{Getal05}).

In this paper the entrainment matrix of
a neutron-proton mixture
is derived for non-zero temperatures.
Calculations are performed
in the frame of the Landau Fermi-liquid
theory, generalized by Larkin and Migdal \cite{LM63}
and Leggett \cite{L65}
to the case of superfluidity.
For the sake of simplicity
we assume the singlet-state ($^1S_0$)
pairing of nucleons of both species.
In Section 4.1 we will show
how the obtained results
can be extended to the case of triplet-state $^3P_2$
neutron pairing.

\section{A current-free neutron-proton mixture}
\label{2}
Before calculating the matrix $\rho_{\alpha \alpha'}$,
let us consider a neutron-proton mixture in the absence of currents.
A simple generalization of the Hamiltonian
of a superfluid Fermi-liquid, suggested
by Leggett \cite{L65},
to the case of superfluid mixtures, gives
\begin{eqnarray}
{\rm H}- \mu_{\rm n} {\rm N_n} - \mu_{\rm p} {\rm N_p}
&=& {\rm H}_{\rm LF} + {\rm H}_{\rm pairing} \,.
\label{hamilt}
\end{eqnarray}
Here, ${\rm H}$ is the hamiltonian of the system which is the sum
of the standard Fermi-liquid hamiltonian 
${\rm H}_{\rm LF}$ for mixtures and the
pairing hamiltonian ${\rm H}_{\rm pairing}$; ${\rm
N}_{\alpha}$ and $\mu_{\alpha}$ are the number density
operator and the chemical potential 
of nucleon species $\alpha$, respectively.
The expression for ${\rm H}_{\rm LF}$ has the form
(see, e.g., Ref. \cite{S76}):
\begin{eqnarray}
{\rm H}_{\rm LF} &=& \sum_{{\pmb p} \sigma \alpha}
\varepsilon_{0}^{(\alpha)} \left({\pmb p} \right) \,
\left( \apc \ap - \thp \right)
\nonumber \\
&+& \frac{1}{2}  \sum_{{\pmb p} {\pmb p} '  \sigma \sigma ' \alpha \alpha '}
f^{\alpha \alpha'} \left(\vp, \vps \right)
\left( \apc \ap - \thp \right) \left( \apsc \aps - \thps \right) \,.
\label{LF}
\end{eqnarray}
In Eq. (\ref{LF}) the summation is taken over the particle
momenta $\vp$ and $\vps$, as well as over the spin projections
$\sigma$ and $\sigma'$
onto the quantization axis and over the particle species
$\alpha, \alpha' = {\rm n}$ or ${\rm p}$;
$a_{\pmb p}^{(\alpha)} \equiv a_{{\pmb p} \sigma}^{(\alpha)} =
a_{{\pmb p} \uparrow}^{(\alpha)}$ or $a_{{\pmb p} \downarrow}^{(\alpha)}$
is the annihilation operator of a quasiparticle
(not the Bogoliubov excitation!)
of species $\alpha$ in a state (${\pmb p} \sigma$).
We restrict ourselves to a spin-unpolarized nucleon matter.
This allows us to simplify the notations.
For instance, we will drop spin
indices, whenever possible.
Furthermore,
$\thp = \theta \left( \pFa - |\vp| \right)$, where $\theta(x)$
is the step function;
$\varepsilon_{0}^{(\alpha)} \left({\pmb p} \right) = \vFa (|\vp|-\pFa)$,
where $\vFa$ and $\pFa$ are, respectively, 
the Fermi-velocity and Fermi-momentum;
$f^{\alpha \alpha'}({\pmb p}, {\pmb p'})$
is the spin-averaged
Landau quasiparticle interaction
(we disregard the spin-dependence of this interaction
since it does not affect our results);
$f^{\alpha \alpha'}({\pmb p}, {\pmb p'})$
is the (spin-averaged)
second variational derivative of the energy
with respect to the number of particles.
Therefore, it is invariant
under transformations
$\vp \rightleftharpoons \vps$
and $\alpha \rightleftharpoons \alpha'$
(see, e.g., Refs. \cite{LP80,S76}):
\begin{eqnarray}
f^{\alpha \alpha'} \left(\vp, \vps \right) &=&
f^{\alpha' \alpha} \left(\vp, \vps \right)  =
f^{\alpha \alpha'} \left(\vps, \vp \right).
\label{sym1}
\end{eqnarray}
The pairing hamiltonian
can be written as:
\begin{equation}
{\rm H_{\rm pairing}} = \sum_{{\pmb p} {\pmb p}' \alpha}
\mathcal{V}^{(\alpha)} \left({\pmb p}, {\pmb p'} \right) \,\,
a_{{\pmb p}' \uparrow}^{(\alpha) \dagger} \,
a_{-{\pmb p}' \downarrow}^{(\alpha) \dagger} \,
a_{-{\pmb p} \downarrow}^{(\alpha)}  \,
a_{{\pmb p} \uparrow}^{(\alpha)} \,.
\label{sfl}
\end{equation}
Here we assume the following symmetry conditions:
\begin{equation}
\mathcal{V}^{(\alpha)} \left({\pmb p}, {\pmb p'} \right) =
\mathcal{V}^{(\alpha)} \left({\pmb p'}, {\pmb p} \right) =
\mathcal{V}^{(\alpha)} \left({\pmb p}, -{\pmb p'} \right) =
\mathcal{V}^{(\alpha)} \left(-{\pmb p}, {\pmb p'} \right).
\label{sym2}
\end{equation}
If the matrix element 
$\mathcal{V}^{(\alpha)} \left({\pmb p}, {\pmb p'} \right)$
does not satisfy these conditions
(e.g., for singlet-state pairing
of nucleon pairs with
non-zero orbital angular momentum),
it should be symmetrized
in such a way as to obey
Eq. (\ref{sym2}) (see, e.g., Ref. \cite{L75}).
To diagonalize the operator (\ref{hamilt})
we introduce the standard operators
of Bogoliubov excitations $b_{{\pmb p} \sigma}^{(\alpha)}$:
\begin{eqnarray}
a_{{\pmb p} \uparrow}^{(\alpha)} = \uup b_{{\pmb p} \uparrow}^{(\alpha)}
+ \vvp b_{-{\pmb p} \downarrow}^{(\alpha) \dagger},
\label{bp1} \\
a_{{\pmb p} \downarrow}^{(\alpha)} = \uup b_{{\pmb p} \downarrow}^{(\alpha)}
- \vvp b_{-{\pmb p} \uparrow}^{(\alpha) \dagger}.
\label{bp2}
\end{eqnarray}
The parameters $\uup$ and $\vvp$ are 
related by the inversion symmetry and normalization
\begin{equation}
\uup = u_{-{\pmb p}}^{(\alpha)}, \quad
\vvp = v_{-{\pmb p}}^{(\alpha)}, \quad
u_{\pmb p}^{(\alpha) \, 2} + v_{\pmb p}^{(\alpha) \, 2} = 1 \,.
\label{norm}
\end{equation}
Now one can find the eigenvalues of the operator (\ref{hamilt})
\begin{eqnarray}
E-\mu_{\rm n}  n_{\rm n} &-& \mu_{\rm p} n_{\rm p} =
\sum_{{\pmb p} \sigma \alpha}
\varepsilon_0^{(\alpha)} \left( {\pmb p} \right) \,
\left( \np - \thp \right)
\nonumber \\
&+& \frac{1}{2}  \sum_{{\pmb p} {\pmb p} '  \sigma \sigma ' \alpha \alpha '}
f^{\alpha \alpha'} \left(\vp, \vps \right)
\left( \np - \thp \right) \left( \nps - \thps \right)
\nonumber \\
&+& \sum_{{\pmb p} {\pmb p}' \alpha}
\mathcal{V}^{(\alpha)} \left({\pmb p}, {\pmb p'} \right) \,\,
\uup \vvp \,
u_{{\pmb p}  '}^{(\alpha)} v_{{\pmb p}  '}^{(\alpha)}\, \,
\left(  1-2 \fp\right) \,  
\left(  1-2 \mathfrak{f}_{{\pmb p}  '}^{(\alpha)} \right),
\label{energy}
\end{eqnarray}
where $\np$ and $\fp$ are the distribution functions of
quasiparticles and Bogoliubov
excitations, respectively, given by
\begin{eqnarray}
\np &=& \,\, \langle | a_{{\pmb p} \uparrow}^{(\alpha) \dagger} a_{{\pmb p}
\uparrow}^{(\alpha)} | \rangle
\,\, = \,\, \langle| a_{{\pmb p} \downarrow}^{(\alpha) \dagger} a_{{\pmb p}
\downarrow}^{(\alpha)} | \rangle
\, = v_{\pmb p}^{(\alpha) \, 2} +
\left( u_{\pmb p}^{(\alpha) \, 2} - v_{\pmb p}^{(\alpha) \, 2}\right) \, \fp ,
\label{np} \\
\fp &=& \,\, \langle | b_{{\pmb p} \uparrow}^{(\alpha) \dagger} b_{{\pmb p}
\uparrow}^{(\alpha)} |\rangle
\,\, = \,\,
\langle | b_{{\pmb p} \downarrow}^{(\alpha) \dagger} b_{{\pmb p}
\downarrow}^{(\alpha)} | \rangle.
\label{fp}
\end{eqnarray}
The entropy of the system is given by the usual
combinatorial expression:
\begin{equation}
S = - \sum_{{\pmb p} \sigma \alpha}
\left[ \left( 1-\fp \right) \ln \left( 1-\fp\right) + \fp  \ln \fp
\right] \, .
\label{entropy}
\end{equation}
Minimizing the thermodynamical potential
$F =E-\mu_{\rm n}  n_{\rm n} - \mu_{\rm p} n_{\rm p} - T S$
with respect to unknown functions $\fp$ and $\uup$
($\vvp$ can be expressed through $\uup$ using Eq. (\ref{norm})),
one finds
\begin{eqnarray}
\fp &=& \frac{1}{1 + {\rm e}^{E_{\pmb p}^{(\alpha)}/T}}, \quad
E_{\pmb p}^{(\alpha)} = \sqrt{\varepsilon^{(\alpha) 2} \left({\pmb p} \right)
+ \Delta_{\pmb p}^{(\alpha) 2}},
\label{distribution} \\
u_{{\pmb p}}^{(\alpha) \, 2} &=& \frac{1}{2} \,
\left( 1 + \frac{\varepsilon^{(\alpha)}(\vp)}{E_{\pmb p}^{(\alpha)}}
\right) \, .
\label{up}
\end{eqnarray}
Here and below we use the system
of units in which $\hbar=\kB=V=1$, where $\hbar$ is the Planck constant,
$\kB$ is the Boltzmann constant, and $V$ is the normalization volume.
In Eqs. (\ref{distribution}) -- (\ref{up}) $E_{\pmb p}^{(\alpha)}$
is the energy of a Bogoliubov excitation with momentum $\vp$.
The superfluid gap $\Delta_{\pmb p}^{(\alpha)}$ of
nucleon species $\alpha$ can be determined from the equation
\begin{equation}
\Delta_{\pmb p}^{(\alpha)} = - \sum_{{\pmb p}'}
\mathcal{V}^{(\alpha)} \left({\pmb p}, {\pmb p'} \right)
u_{{\pmb p}'}^{(\alpha)} v_{{\pmb p}'}^{(\alpha)}
\left( 1- 2  \mathfrak{f}_{{\pmb p}  '}^{(\alpha)}\right) \,.
\label{gap}
\end{equation}
Finally, $\varepsilon^{(\alpha)} \left({\pmb p} \right)$
is the quantity which formally coincides with
the energy of quasiparticle species $\alpha$
in the mixture of non-superfluid Fermi-liquids,
\begin{equation}
\varepsilon^{(\alpha)} \left({\pmb p} \right)=
\varepsilon_0^{(\alpha)} \left({\pmb p} \right) + \sum_{{\pmb p}' \sigma'
\alpha'}
f^{\alpha \alpha'} \left(\vp, \vps \right) \,
\left( \nps - \thps \right) \, .
\label{localenergy}
\end{equation}
In Eq. (\ref{localenergy}) the quasiparticle distribution
function $\np$ is determined by Eq. ($\ref{np}$)
with $\fp$ and $\uup$
taken from Eqs. (\ref{distribution})--(\ref{up}).
The first term in the right-hand side of Eq. (\ref{localenergy})
can be estimated as:
$\varepsilon_0^{(\alpha)} \left({\pmb p} \right)
\sim \left(T+\Delta^{(\alpha)} \right)$,
where $\Delta^{(\alpha)}$ is a typical value of the gap.
In thermodynamic equilibrium, the second term in
Eq. (\ref{localenergy}) is much
smaller than the first one
because for any function $f(p)$, 
smooth in the vicinity of a Fermi surface,
one has the estimate:
\begin{equation}
\int_{0}^{\infty} \, f(p) \, p^2 \, (\np - \thp) \, \dd p \,\, \sim
f(\pFa) \,\,  n_{\alpha} \, \left( [T/\mu_\alpha]^2 +
[\Delta^{(\alpha)}/\mu_\alpha]^2
\right) \, ,
\label{example}
\end{equation}
Thus, since
$\left(T + \Delta^{(\alpha)} \right)/\mu_{\alpha} \ll 1$,
the second term in Eq. (\ref{localenergy}) can be neglected.

\section{A neutron-proton mixture with superfluid currents}
\label{3}
\subsection{General consideration}
In a system with superfluid currents
the plane-wave states of nucleons
$({\pmb p}+{\pmb Q}_{\alpha}, \uparrow)$
and $(-{\pmb p}+{\pmb Q}_{\alpha}, \downarrow)$
are paired
(note, that we consider singlet-state pairing
of both species and assume
${\pmb Q}_{\alpha} \equiv m_{\alpha} {\pmb V}_{\alpha s} \ll \pFa$).
In this case the pairing hamiltonian
should be written as
(see, e.g., Refs. \cite{F72,G66})
\begin{equation}
{\rm H_{\rm pairing}}({\pmb Q}_{\alpha}) = \sum_{{\pmb p} {\pmb p}' \alpha}
\mathcal{V}^{(\alpha)}_{{\pmb Q}_{\alpha}} \left({\pmb p}, {\pmb p'} \right) \,
a_{{\pmb p}' + {\pmb Q}_{\alpha} \uparrow}^{(\alpha) \dagger}
a_{-{\pmb p}' + {\pmb Q}_{\alpha} \downarrow}^{(\alpha) \dagger} \,
a_{-{\pmb p} + {\pmb Q}_{\alpha} \downarrow}^{(\alpha) }
a_{{\pmb p} + {\pmb Q}_{\alpha} \uparrow}^{(\alpha) }.
\label{BCScurrent}
\end{equation}
Here $\mathcal{V}^{(\alpha)}_{{\pmb Q}_{\alpha}} \left({\pmb p}, {\pmb p'} \right)$
is the matrix element for the scattering of a pair
of quasiparticles species ${\alpha}$ from states
$({\pmb p}+{\pmb Q}_{\alpha}, \uparrow)$, $(-{\pmb p}+{\pmb Q}_{\alpha},
\downarrow)$ to states $({\pmb p'}+{\pmb Q}_{\alpha}, \uparrow)$,
$(-{\pmb p'}+{\pmb Q}_{\alpha}, \downarrow)$.
%
Due to the rotational invariance, the expansion of
$\mathcal{V}^{(\alpha)}_{{\pmb Q}_{\alpha}} \left({\pmb p}, {\pmb p'} \right)$
in powers of ${\pmb Q}_{\alpha}$ will contain the terms 
$ \sim Q_{\alpha}^2$ and higher.
Therefore, as we
will work in the linear approximation in
${\pmb Q}_{\alpha}$, we will neglect the dependence of
$\mathcal{V}^{(\alpha)}_{{\pmb Q}_{\alpha}} \left({\pmb p}, {\pmb p'} \right)$
on momentum ${\pmb Q}_{\alpha}$ and put 
$\mathcal{V}^{(\alpha)}_{{\pmb Q}_{\alpha}} \left({\pmb p}, {\pmb p'}
\right) \approx
\mathcal{V}^{(\alpha)} \left({\pmb p}, {\pmb p'} \right)$.

Expressing the quasiparticle operators in terms of Bogoliubov
excitation operators
\begin{eqnarray}
a_{{\pmb p} + {\pmb Q}_{\alpha} \uparrow}^{(\alpha)}
&=& \uuup b_{{\pmb p} + {\pmb Q}_{\alpha} \uparrow}^{(\alpha)}
+ \vvvp b_{-{\pmb p} + {\pmb Q}_{\alpha} \downarrow}^{(\alpha) \dagger},
\label{bpQ1} \\
a_{{\pmb p} + {\pmb Q}_{\alpha} \downarrow}^{(\alpha)}
&=& \uuup b_{{\pmb p} + {\pmb Q}_{\alpha} \downarrow}^{(\alpha)}
- \vvvp b_{-{\pmb p} + {\pmb Q}_{\alpha} \uparrow}^{(\alpha) \dagger},
\label{bpQ2}
\end{eqnarray}
where $\uuup$ and $\vvvp$
satisfy the equalities
\begin{equation}
\uuup = U_{-{\pmb p}}^{(\alpha)}, \quad
\vvvp = V_{-{\pmb p}}^{(\alpha)}, \quad
U_{\pmb p}^{(\alpha) \, 2} + V_{\pmb p}^{(\alpha) \, 2} = 1,
\label{norm2}
\end{equation}
similar to (\ref{norm}),
one obtains the expression for the energy density:
\begin{gather}
E-\mu_{\rm n}  n_{\rm n} - \mu_{\rm p}  n_{\rm p} =
\sum_{{\pmb p} \sigma \alpha}
\varepsilon_{0}^{(\alpha)}\left({\pmb p}+{\pmb Q}_{\alpha} \right) \,
\left( \mN_{{\pmb p}+ {\pmb Q}_{\alpha}}^{(\alpha)} - \thpQ \right)
\nonumber \\
+ \frac{1}{2}  \sum_{{\pmb p} {\pmb p} '  \sigma \sigma ' \alpha \alpha '}
f^{\alpha \alpha'} \left(\vp + \vQa, \vps + {\pmb Q}_{\alpha'} \right)
\left( \mN_{{\pmb p}+ {\pmb Q}_{\alpha}}^{(\alpha)} - \thpQ \right)
\left( \mN_{{\pmb p}'+ {\pmb Q}_{\alpha'}}^{(\alpha')} - \thpsQ \right)
\nonumber \\
+ \sum_{{\pmb p} {\pmb p}' \alpha}
\mathcal{V}^{(\alpha)} \left({\pmb p}, {\pmb p'} \right) \,\,
\uuup \vvvp \,
U_{{\pmb p}  '}^{(\alpha)} V_{{\pmb p}  '}^{(\alpha)}\, \,
\nonumber \\
\times \,\, \left(  1- \mF_{{\pmb p}+ {\pmb Q}_{\alpha}}^{(\alpha)}
- \mF_{-{\pmb p}+ {\pmb Q}_{\alpha}}^{(\alpha)}\right) \,
\left(  1-\mF_{{\pmb p}'+ {\pmb Q}_{\alpha}}^{(\alpha)}
-\mF_{-{\pmb p}'+ {\pmb Q}_{\alpha}}^{(\alpha)}\right).
\label{energy1}
\end{gather}
Here $\mN_{{\pmb p}+ {\pmb Q}_{\alpha}}^{(\alpha)}$ and
$\mF_{{\pmb p}+ {\pmb Q}_{\alpha}}^{(\alpha)}$
are the distribution functions of quasiparticles and Bogoliubov excitations
with momentum (${\pmb p}+ {\pmb Q}_{\alpha}$), respectively:
\begin{eqnarray}
\mN_{{\pmb p}+ {\pmb Q}_{\alpha}}^{(\alpha)} &=& \,\,
\langle | a_{{\pmb p} +{\pmb Q}_{\alpha} \uparrow}^{(\alpha) \dagger}
a_{{\pmb p} +{\pmb Q}_{\alpha} \uparrow}^{(\alpha)} |\rangle
\,\, = \,\,
\langle | a_{{\pmb p} +{\pmb Q}_{\alpha}\downarrow}^{(\alpha) \dagger}
a_{{\pmb p}+{\pmb Q}_{\alpha} \downarrow}^{(\alpha)} |\rangle
\nonumber \\
&=& V_{\pmb p}^{(\alpha) \, 2} +
 U_{\pmb p}^{(\alpha) \, 2} \, \mF_{{\pmb p}+ {\pmb Q}_{\alpha}}^{(\alpha)}
- V_{\pmb p}^{(\alpha) \, 2} \, \mF_{-{\pmb p}+ {\pmb Q}_{\alpha}}^{(\alpha)},
\label{np1} \\
\mF_{{\pmb p}+ {\pmb Q}_{\alpha}}^{(\alpha)} &=&
\langle | b_{{\pmb p} +{\pmb Q}_{\alpha} \uparrow}^{(\alpha) \dagger}
b_{{\pmb p} +{\pmb Q}_{\alpha} \uparrow}^{(\alpha)} |\rangle
\,\, = \,\,
\langle | b_{{\pmb p} +{\pmb Q}_{\alpha}\downarrow}^{(\alpha) \dagger}
b_{{\pmb p}+{\pmb Q}_{\alpha} \downarrow}^{(\alpha)} | \rangle.
\label{fp1}
\end{eqnarray}
The entropy of the system is still given by Eq. (\ref{entropy})
with the distribution function $\fp$ replaced by $\mF_{{\pmb p}+ {\pmb
Q}_{\alpha}}^{(\alpha)}$.
The minimization of the thermodynamical potential
$F =E-\mu_{\rm n}  n_{\rm n}
- \mu_{\rm p} n_{\rm p} - T S$
with respect to $\mF_{{\pmb p}+ {\pmb Q}_{\alpha}}^{(\alpha)}$
and $\uuup$ yields
\begin{eqnarray}
\mF_{{\pmb p}+ {\pmb Q}_{\alpha}}^{(\alpha)} &=&
\frac{1}{ 1 + {\rm e}^{\mathfrak{E}_{{\pmb p}+ {\pmb Q}_{\alpha}}^{(\alpha)}/T}}
\,,
\label{Fp} \\
\mathfrak{E}_{{\pmb p}+ {\pmb Q}_{\alpha}}^{(\alpha)}
&=& \frac{1}{2} \, \left( H_{{\pmb p}+ {\pmb Q}_{\alpha}}^{(\alpha)}
-H_{-{\pmb p}+ {\pmb Q}_{\alpha}}^{(\alpha)}
\right) \, +
\sqrt{\, \frac{1}{4} \, \left( H_{{\pmb p}+ {\pmb Q}_{\alpha}}^{(\alpha)}
+H_{-{\pmb p}+ {\pmb Q}_{\alpha}}^{(\alpha)}\right)^2
+ \mathcal{D}_{\pmb p}^{(\alpha) 2} }\, ,
\label{distribution1} \\
U_{{\pmb p}}^{(\alpha) \, 2} &=& \frac{1}{2} \,
\left( 1 + { H_{{\pmb p}+ {\pmb Q}_{\alpha}}^{(\alpha)}
+H_{-{\pmb p}+ {\pmb Q}_{\alpha}}^{(\alpha)}
\over 2 \mathfrak{E}_{{\pmb p}+ {\pmb Q}_{\alpha}}^{(\alpha)}
+ H_{-{\pmb p}+ {\pmb Q}_{\alpha}}^{(\alpha)}
- H_{{\pmb p}+ {\pmb Q}_{\alpha}}^{(\alpha)}} \right) \,.
\label{upQ}
\end{eqnarray}
In Eqs. (\ref{Fp})--(\ref{upQ})
$\mathfrak{E}_{{\pmb p}+ {\pmb Q}_{\alpha}}^{(\alpha)}$
is the energy of a Bogoliubov excitation
with momentum (${\pmb p}+ {\pmb Q}_{\alpha}$).
The stability of the system implies
$\mathfrak{E}_{{\pmb p}
+ {\pmb Q}_{\alpha}}^{(\alpha)} \geq 0$.
Furthermore, $\mathcal{D}_{\pmb p}^{(\alpha)}$ is the superfluid gap
which can be found from the equation:
\begin{equation}
\mathcal{D}_{\pmb p}^{(\alpha)} = - \sum_{{\pmb p}'} \,
\mathcal{V}^{(\alpha)} \left({\pmb p}, {\pmb p'} \right) \,
U_{{\pmb p}'}^{(\alpha)} V_{{\pmb p}'}^{(\alpha)}
\, \left(  1-\mF_{{\pmb p}'+ {\pmb Q}_{\alpha}}^{(\alpha)}
-\mF_{-{\pmb p}'+ {\pmb Q}_{\alpha}}^{(\alpha)}\right).
\label{gap1}
\end{equation}
Finally, $H_{{\pmb p}+ {\pmb Q}_{\alpha}}^{(\alpha)}$ is the quantity
which formally coincides with the energy of
a quasiparticle with momentum
(${\pmb p}+ {\pmb Q}_{\alpha}$) in the mixture of non-superfluid
Fermi-liquids,
\begin{equation}
H_{{\pmb p}+ {\pmb Q}_{\alpha}}^{(\alpha)} =
\varepsilon_0^{(\alpha)}  \left({\pmb p}+{\pmb Q}_{\alpha} \right)
+ \sum_{{\pmb p}' \sigma' \alpha'}
f^{\alpha \alpha'} \left(\vp+\vQa, \vps+{\pmb Q}_{\alpha '} \right) \,
\left(  \mN_{{\pmb p}' + {\pmb Q}_{\alpha '}}^{(\alpha')} - \thpsQ \right).
\label{localenergy2}
\end{equation}
The quasiparticle distribution function
$\mN_{{\pmb p}+ {\pmb Q}_{\alpha}}^{(\alpha)}$
is defined by Eq. (\ref{np1}).
Since $Q_{\alpha} \ll \pFa$, one can expand
$H_{{\pmb p}+ {\pmb Q}_{\alpha}}^{(\alpha)}$ in terms of
$Q_{\alpha'}$ ($\alpha' = {\rm n}, {\rm p})$ and write
\begin{equation}
H_{{\pmb p}+{\pmb Q}_{\alpha}}^{(\alpha)} = \varepsilon^{(\alpha)}({\pmb p}) +
\Delta H_{\pmb p}^{(\alpha)} \,.
\label{expand1}
\end{equation}
In the case of singlet-state nucleon pairing there are only three
vectors $\vp, \,  {\pmb Q}_{\rm n}$ and  ${\pmb Q}_{\rm p}$
which can form the scalar $\Delta H_{\pmb p}^{(\alpha)}$.
Neglecting all terms which are quadratic and
higher order in $Q_{\alpha}/\pFa$, one can write
\begin{equation}
\Delta H_{\pmb p}^{(\alpha)} = \sum_{\alpha '} \, \gamma_{\alpha \alpha'}(p) \,
\,\,
{ \vp \, {\pmb Q}_{\alpha'} \over m_{\alpha'}} \,,
\label{dH}
\end{equation}
where $\gamma_{\alpha \alpha'}(p)$ is the matrix
to be derived in the next section on the Fermi surface of
particle species $\alpha$ (at $p = \pFa$).
Taking into account Eqs. (\ref{sym2}),
(\ref{distribution1})--(\ref{gap1}), and (\ref{expand1})--(\ref{dH})
and neglecting the terms $\sim Q_{\alpha}^2$, one has
\begin{equation}
\mathcal{D}_{\pmb p}^{(\alpha)} = {\Delta}_{\pmb p}^{(\alpha)}, \quad
U_{\pmb p}^{(\alpha)} = u_{\pmb p}^{(\alpha)}, \quad
V_{\pmb p}^{(\alpha)}=v_{\pmb p}^{(\alpha)}.
\label{Equal}
\end{equation}
Now the energy of Bogoliubov
excitations $\mathfrak{E}_{{\pmb p}+ {\pmb Q}_{\alpha}}^{(\alpha)}$,
as well as the distribution functions of quasiparticles
and Bogoliubov excitations can be expanded
in analogy with Eq. (\ref{expand1}) as
\begin{eqnarray}
\mathfrak{E}_{{\pmb p}+ {\pmb Q}_{\alpha}}^{(\alpha)}
&=& E_{\pmb p}^{(\alpha)} + \Delta H_{\pmb p}^{(\alpha)} \,,
\label{expand3} \\
\mF_{{\pmb p}+{\pmb Q}_{\alpha}}^{(\alpha)} &=&
\fp + {\partial \fp \over \partial E_{\pmb p}^{(\alpha)}} \,\, \Delta H_{\pmb
p}^{(\alpha)}, \quad
\mN_{{\pmb p}+{\pmb Q}_{\alpha}}^{(\alpha)} =
\np + {\partial \fp \over \partial E_{\pmb p}^{(\alpha)}} \,\, \Delta H_{\pmb
p}^{(\alpha)} \,.
\label{expand2}
\end{eqnarray}
%
\subsection{The calculation of matrix $\gamma_{\alpha \alpha'}$}
To calculate the matrix $\gamma_{\alpha \alpha'}(\pFa)$
we will make use of Eq. (\ref{localenergy2}).
Restricting ourselves to the terms
linear in ${\pmb Q}_{\alpha}$,
we expand all functions
in Eq. (\ref{localenergy2})
using the formulas (\ref{expand1}) and (\ref{expand2}).
Then, taking into account 
Eqs. (\ref{localenergy}) and (\ref{example}), and
neglecting all terms in Eq. (\ref{localenergy2}) which
depend on $(\nps-\thps)$, we obtain
\begin{equation}
\Delta H_{\pmb p}^{(\alpha)} = { {\pmb p \, {\pmb Q}_{\alpha}} \over
m_\alpha^\ast}
+ \sum_{{\pmb p}' \sigma' \alpha'} \,
f^{\alpha \alpha'} \left( {\pmb p}, {\pmb p}' \right) \,
\left\{
{\partial \mathfrak{f}_{{\pmb p}'}^{(\alpha')} \over \partial E_{{\pmb p}'}^{(\alpha')} }
\,\,
\Delta H_{{\pmb p}'}^{(\alpha')} -
{\partial \theta_{{\pmb p}'}^{(\alpha')}  \over \partial {\pmb p}'} \, {\pmb
Q}_{\alpha'}
\right\} \, .
\label{Eq}
\end{equation}
Let us calculate the sum in this equation.
The main contribution to
the sum comes
from a narrow region of
$|{\pmb p}'| \sim \pFas$ since
the function in the curly
brackets is essentially non-zero
only close to the Fermi surface
of particle species $\alpha'$.
Furthermore, in a smooth function
$f^{\alpha \alpha'} \left( {\pmb p}, {\pmb p}' \right)$
we replace $|{\pmb p}|$ and $|{\pmb p}'|$
by $\pFa$ and $\pFas$, respectively,
and expand it into Legendre polynomials $P_l(\cos \theta)$:
\begin{equation}
f^{\alpha \alpha'} \left( {\pmb p}, {\pmb p}' \right) = \sum_{l} \,
f^{\alpha \alpha'}_l \, P_l (\cos \theta),
\label{legendre}
\end{equation}
where $\theta$ is the angle between
${\pmb p}$ and ${\pmb p}'$.
Using isotropy of the gaps 
$\Delta_{\pmb p}^{(\alpha)}$ ($\alpha = {\rm n, p}$)
and Eq. (\ref{dH}) we obtain
\begin{eqnarray}
\sum_{{\pmb p}' \sigma'} \,
f^{\alpha \alpha'} \left( {\pmb p}, {\pmb p}' \right) \,
{\partial \mathfrak{f}_{{\pmb p}'}^{(\alpha')} \over \partial E_{{\pmb p}'}^{(\alpha')} }
\,\,
\Delta H_{{\pmb p}'}^{(\alpha')} &=&  - \, { f^{\alpha \alpha'}_1 N_{0 \alpha'}
\over 3} \,\,
{\pF_{\alpha'} \over \pF_\alpha} \,\, \Phi_{\alpha'} \,
\Delta H_{{\pmb p}}^{(\alpha')} \, ,
\label{intF} \\
\sum_{{\pmb p}' \sigma'} \,
f^{\alpha \alpha'} \left( {\pmb p}, {\pmb p}' \right) \,
{\partial \theta_{{\pmb p}'}^{(\alpha')}  \over \partial {\pmb p}'} \, {\pmb
Q}_{\alpha'}
&=& - \, { f^{\alpha \alpha'}_1 N_{0 \alpha'} \over 3} \,\,
{\pF_{\alpha'} \over \pF_\alpha} \,\, {{\pmb p} \, {\pmb Q}_{\alpha'} \over
m^{\ast}_{\alpha'}} \, .
\label{intQ}
\end{eqnarray}
Here $N_{0 \alpha} = m_{\alpha}^\ast \, \pFa/\pi^2$;
$m_{\alpha}^\ast = \pFa/\vFa$ is the effective mass of particle
species $\alpha$.
The function $\Phi_{\alpha}$, which is given by
\begin{equation}
\Phi_{\alpha} = - { 1 \over N_{0 \alpha}} \, \sum_{{\pmb p} \sigma} \,
{\partial \fp \over \partial E_{\pmb p}^{(\alpha)}} \,,
\label{N0}
\end{equation}
was calculated numerically and approximated
by Gnedin and Yakovlev \cite{GY95}.
Their fit of $\Phi_\alpha$ is given in Appendix A.
When calculating $\Phi_\alpha$, the authors
adopted the standard
approximation in which the dependence of the gap on
the absolute value of particle momentum is neglected
(consequently, in the isotropic case the gap is a function
of temperature only),
$\Delta^{(\alpha)}_{\pmb p} \approx \Delta^{(\alpha)}(|{\pmb p}|=\pFa) \equiv
\Delta^{(\alpha)}(T)$.
In this paper we also use this approximation.

Now, writing Eq. (\ref{Eq}) for neutrons
($\alpha={\rm n}$) and for protons ($\alpha={\rm p}$),
taking into account Eqs. (\ref{dH}) and (\ref{intF})--(\ref{intQ}),
and equating prefactors at the same ${\pmb Q}_{\alpha}$
in its left- and right-hand sides,
we arrive at the set of
linear equations for
the matrix $\gamma_{\alpha \alpha'}$.
Its solution has the form:
\begin{eqnarray}
\gamma_{\alpha \alpha}(\pF_{\alpha}) &=&
{m_{\alpha} \over m_{\alpha}^{\ast}} \,\, {1 \over S} \,\,
\left\{
\left( 1 + {F_1^{\alpha \alpha} \over 3} \right) \,
\left( 1 + {F_1^{\beta \beta} \over 3} \, \Phi_\beta \right)
- \left( {F_1^{\alpha \beta} \over 3 }\right)^2 \, \Phi_{\beta} \right\} \,,
\label{solutions1} \\
\gamma_{\alpha \beta}(\pF_{\alpha}) &=&
{1 \over 3} \,\,
{ m_{\beta} \over \sqrt{m_{\alpha}^{\ast} \, m_{\beta}^{\ast}}  } \,\,
{1 \over S} \,\,
\left( {\pF_{\beta} \over \pF_{\alpha} } \right)^{3/2}  \, \,
F_1^{\alpha \beta} \, (1-\Phi_{\beta}) \, ,
\label{solutions2} \\
S &\equiv& \left( 1 + {F_1^{\alpha \alpha} \over 3} \, \Phi_\alpha \right) \,
\left( 1 + {F_1^{\beta \beta} \over 3} \, \Phi_\beta \right)
- \left( {F_1^{\alpha \beta} \over 3 }\right)^2 \, \Phi_\alpha \Phi_{\beta} \,,
\label{S} \\
F_1^{\alpha \beta} &\equiv& f^{\alpha \beta}_1 \, 
\sqrt{N_{0 \alpha} N_{0 \beta}} \,.
\label{Landau}
\end{eqnarray}
Here $\alpha \neq \beta$ (thus if, e.g., $\alpha= {\rm n}$
then $\beta= {\rm p}$).
The effective masses $m_{\alpha}^\ast$ are related to the
parameters $f_1^{\alpha \alpha'}$ through the equation
(see Refs. \cite{S76,Betal96}):
\begin{equation}
{m_{\alpha}^\ast \over m_{\alpha}} = 1 + {N_{0 \alpha} \over 3}
\left[ f_1^{\alpha \alpha} + {m_{\beta} \over m_{\alpha}} \,
\left( \pFb \over \pFa \right)^2 \, f_1^{\alpha \beta} \right] \,, \quad \quad
\alpha \neq \beta \,.
\label{effmass}
\end{equation}
%

\section{The entrainment matrix}
\label{4}
\subsection{Superfluid current
and the matrix $\rho_{\alpha \alpha'}$ in different limiting cases}
Eqs. (\ref{expand1}) and (\ref{expand3})--(\ref{expand2})
allow one to calculate the entrainment matrix $\rho_{\alpha \alpha'}$
and to express it in terms of
$\gamma_{\alpha \alpha'}(\pFa)$.
This can be done, for instance,
with the aid of Eq. (\ref{AB})
and the energy density given by Eq. (\ref{energy1}).
It is easier, however, to obtain
$\rho_{\alpha \alpha'}$
by calculating the mass current density ${\pmb J}_{\alpha}$
of quasiparticles
(we remind that ${\pmb V}_{\rm qp}=0$ in a chosen coordinate frame).
When doing so we take account of the fact that the expression
for the mass current density of non-superfluid Fermi-liquid
can be applied to the superfluid state as well
(see, e.g., Refs. \cite{L65,Betal96}).
In our case this means that
the expression for ${\pmb J}_{\alpha}$
has the form
(see, e.g., Refs. \cite{LP80,BP91}):
\begin{equation}
{\pmb J}_{\alpha} = \sum_{{\pmb p \sigma}} \, m_{\alpha} \,
{\partial H_{{\pmb p}+{\pmb Q}_{\alpha}}^{(\alpha)} \over \partial {\pmb p}}
\,\, \mN_{{\pmb p}+{\pmb Q}_{\alpha}}^{(\alpha)}.
\label{J}
\end{equation}
Substituting Eqs. (\ref{expand1}) and (\ref{expand2})
into Eq. (\ref{J}) and performing a simple integration,
we obtain the formulas, similar to Eqs. (\ref{Jn})--(\ref{Jp}),
with the entrainment matrix equal to ($\alpha, \alpha' = {\rm n, p}$):
\begin{equation}
{\rho}_{\alpha \alpha'} = \rho_{\alpha} \, 
\gamma_{\alpha \alpha'}(\pF_\alpha) \,
\left( 1-\Phi_{\alpha} \right) .
\label{matrix} 
\end{equation}
From Eq. (\ref{solutions2}) it follows that the entrainment
matrix is indeed symmetric in accordance with Eq. (\ref{AB}),
$\rho_{\alpha \alpha'} = \rho_{\alpha' \alpha}$.
At $T=T_{c \alpha}$
(where $T_{c \alpha}$ is the critical temperature of
quasiparticle species $\alpha$)
one has $\Phi_{\alpha}=1$,
and consequently, $\rho_{\alpha \alpha'} = 0$
and ${\pmb J}_{\alpha} =0$.

Let us analyze the matrix $\rho_{\alpha \alpha'}$
in different limiting cases.

I. Let $F^{\rm np}_1 = 0$, i.e., the
interaction between neutrons and protons is absent.
Then each component $\alpha$ of the mixture
can be treated as an independent superfluid Fermi-liquid
while the matrix $\rho_{\alpha \alpha'}$ is diagonal.
Eqs. (\ref{matrix}) and
(\ref{solutions1})--(\ref{Landau}) yield
\begin{equation}
\rho_{\alpha \alpha} = \frac{\rho_{\alpha} \, (1-\Phi_{\alpha})}
{1+ \Phi_\alpha \,F_{1}^{\alpha \alpha}/3}\, , \quad
\rho_{\rm np} = \rho_{\rm pn}=0 \,.
\label{Leggett}
\end{equation}
The diagonal elements $\rho_{\alpha \alpha}$
coincide with the well-known expression for the
superfluid density of one component Fermi-liquid
(see, e.g., Refs. \cite{L65,BN69}).

II. Let the temperature of the mixture be equal to zero.
Then $\Phi_{\alpha}=0$ and
the entrainment matrix can be rewritten in the form
\begin{eqnarray}
\rho_{\alpha \alpha} &=& \rho_{\alpha} \, \frac{m_{\alpha}}{m_{\alpha}^\ast}
\,\,
\left( 1+ \frac{F_{1}^{\alpha \alpha}}{3} \right) \, ,
\label{Borumand1} \\
\rho_{\rm np} &=& \rho_{\rm pn} = \frac{\pFn^2 \, \pFp^2}{9 \, \pi^4}   \,
\, m_{\rm n} m_{\rm p} \,\, f_{1}^{\rm np} \, ,
\label{Borumand2}
\end{eqnarray}
in agreement with the
results of Borumand et al. \cite{Betal96}.

III. Finally, let us suppose that
the only one component $\alpha$
of the mixture is superfluid.
In this case $\Phi_{\beta}=1$ ($\beta \neq \alpha$)
and we have:
\begin{eqnarray}
\rho_{\alpha \alpha} &=& \rho_{\alpha} \, (1- \Phi_{\alpha}) \,\,
{m_{\alpha} \over m_{\alpha}^{\ast}} \,\,
{\left( 1 + F_1^{\alpha \alpha} /3 \right) \,
\left( 1 + F_1^{\beta \beta} /3 \right)
- \left( F_1^{\alpha \beta} /3 \right)^2
\over \left( 1 + \Phi_\alpha \, F_1^{\alpha \alpha}/3 \right) \,
\left( 1 + F_1^{\beta \beta} /3 \right)
- \left( F_1^{\alpha \beta} /3 \right)^2 \, \Phi_\alpha } \,,
\label{asym3} \\
\rho_{\beta \beta} &=& \rho_{\rm np} = \rho_{\rm pn} = 0\, .
\label{asym33}
\end{eqnarray}

We have calculated the entrainment matrix
of a neutron-proton mixture at non-zero
temperatures assuming singlet-state pairing of
nucleons of both species and isotropic gaps.
However, in reality neutron pairing occurs
in the triplet $^3P_2$ state
of nucleon pair with an anisotropic gap.
In this case the expansion (\ref{dH}) becomes invalid
because the momentum ${\pmb p}$
is no longer the only vector which characterizes
the system in the absence of superfluid currents;
the quantization axis specifies an additional direction.
Therefore, Eq. (\ref{dH}) should be replaced by:
\begin{equation}
\Delta H_{\pmb p}^{(\alpha)} = \sum_{\alpha '} \,
\,\,
{ {\pmb G}_{\alpha \alpha'} \, {\pmb Q}_{\alpha'} \over m_{\alpha'}} \,,
\label{dH1}
\end{equation}
where ${\pmb G}_{\alpha \alpha'}$ is a
matrix composed of vectors.
It can be, in principle, obtained
from Eq. (\ref{localenergy2})
in a manner similar to the derivation of the matrix
$\gamma_{\alpha \alpha'}$.
As follows from Eq. (\ref{J}), components of
the entrainment matrix $\rho_{\alpha \alpha'}$ will be
tensors (rather than scalars as in the isotropic case).
These tensors will be expressed in terms of the Landau
parameters $F_{l}^{\alpha \alpha'}$, with $l \geq 1$.
The quantities $F_{l}^{\alpha \alpha'}$ at $l \geq 2$
are not known for neutron star matter and are not
necessarily small.
Thus, a strict calculation
of the entrainment matrix in the case of triplet-state
neutron pairing is rather complicated
(the analogous problem in deriving the superfluid density
was discussed in details by Leggett \cite{L75} in the context of
the anisotropic phase of helium-3).

To proceed further we follow Baiko et al. \cite{Betal01}
and assume that the neutron star matter
can be treated as a collection of
microscopic domains with arbitrary
orientations of the quantization axis.
Then the entrainment matrix, being averaged over the domains,
will be ``isotropic''
(its elements will be scalars).
Thus, one can use Eq. (\ref{matrix})
for the averaged entrainment matrix by
introducing an {\it effective} isotropic gap
which we choose according to Baiko et al. \cite{Betal01} as
\begin{equation}
\Delta^{(\rm n)}_{\rm eff}(T)= \min
\left\{ \Delta^{(\rm n)}(|{\pmb p}|=\pFn) \right\} \,.
\label{effGAP}
\end{equation}
Here $\Delta^{(\rm n)}_{\rm eff}(T)$
is obtained as the minimum of the angle-dependent
gap $\Delta^{(\rm n)}_{\pmb p}$ on the neutron Fermi surface.
The use of Eq. (\ref{matrix}) with the effective gap
$\Delta^{(\rm n)}_{\rm eff}(T)$ allows one to obtain
qualitatively correct results for the matrix
$\rho_{\alpha \alpha'}$
in the case of triplet-state neutron pairing.
The fit of $\Delta^{(\rm n)}_{\rm eff}(T)$
for the case of $^3P_2$
neutron pairing with $m_{J}=0$ was obtained
by Yakovlev and Levenfish \cite{YL95}
and is given in Appendix A.
\subsection{Landau parameters}
In order to find the matrix $\rho_{\alpha \alpha'}$
it is necessary to know the Landau parameters
$F_1^{\alpha \alpha'}$
(and hence the quantities
$f_1^{\alpha \alpha'}$, see Eq. (\ref{Landau}))
for an asymmetric nuclear matter.
In general, the quantity
$f_l^{\alpha \alpha'}$ is
a function of the baryon number density $n_{\rm b}$
and the asymmetry parameter $\delta=(n_{\rm n}-n_{\rm p})/n_{\rm b}$.
Due to the charge symmetry of strong interactions
one can write
\begin{eqnarray}
f_l^{\rm nn}(n_{\rm b}, \delta) &=& f_l^{\rm pp}(n_{\rm b}, -\delta)
\,,
\label{faa} \\
f_l^{\rm np}(n_{\rm b}, \delta) &=& f_l^{\rm np}(n_{\rm b}, -\delta)
\,.
\label{fnp}
\end{eqnarray}
The function $f_{1}^{\alpha \alpha'}(n_{\rm b}, \delta)$
can be expanded in powers of $\delta\leq 1$.
Neglecting all terms quadratic
and higher order in $\delta$,
Eqs. (\ref{faa})--(\ref{fnp}) yield
\begin{eqnarray}
f_1^{\rm nn} &=& a(n_{\rm b}) + \delta\, b(n_{\rm b})
+ O(\delta^2) \,,
\label{f1nn} \\
f_1^{\rm pp} &=& a(n_{\rm b}) - \delta \, b(n_{\rm b})
+O(\delta^2) \,,
\label{f1pp} \\
f_1^{\rm np} &=& c(n_{\rm b})
+ O(\delta^2) \,.
\label{f1np}
\end{eqnarray}
This approximation was proposed by Haensel \cite{H78}.
To calculate the functions $a(n_{\rm b}), \,\, b(n_{\rm b})$
and $c(n_{\rm b})$ we need to know the dependence
$f_1^{\alpha \alpha'}(n_{\rm b})$ for any two values
of the asymmetry parameter
$\delta=\delta_1$ and $\delta=\delta_2$
(while the knowledge of this dependence
for more than two values of
$\delta$ would enable one to find the terms
non-linear in $\delta$
in the expansions (\ref{f1nn})--(\ref{f1np});
to our best knowledge, these data are unavailable
in the literature).
The  Landau parameters for asymmetric nuclear matter can be
calculated microscopically within a nuclear many-body theory,
starting from the nucleon-nucleon interaction in vacuum.
However, nearly all existing calculations are limited to a
simpler case of a pure neutron matter or symmetric
nuclear matter. In the case of pure neutron matter 
calculations of the $l=1$ Landau parameters 
were performed in some range of
neutron matter density above nuclear density 
(see Refs. \cite{Betal73}--\cite{Setal03}).
In the case of symmetric nuclear matter,
calculations were usually done at normal nuclear density
(see Refs. \cite{S73}--\cite{Betal85})
with a notable exception of the calculation of 
Jackson et al. \cite{Jetal82}. 
We are aware of only one calculations of Landau
parameters in asymmetric nuclear matter in beta equilibrium
with electron gas (simplest model of neutron star matter), by
Shen et al. \cite{Shen03}, who however restricted to $l=0$
parameters, while we need $l=1$ ones. In view of this
situation, we decided to use the density dependent  $l=1$
parameters for symmetric nuclear and pure neutron matter 
calculated by Jackson et al. \cite{Jetal82}
more than two decades ago.

These authors calculated the Landau parameters
for symmetric nuclear matter ($\delta =0$)
and for pure neutron matter ($\delta =1$)
using two model potentials of nucleon-nucleon
interaction: Bethe-Johnson v6 (BJ v6) and Reid v6.
In the case of the symmetric nuclear matter the quantities
$f_1^{\alpha \alpha'}$ can be expressed as
\begin{eqnarray}
f_1^{\rm nn}(n_{\rm b}, 0) &=& f_1^{\rm pp}(n_{\rm b}, 0)
= {F_1(n_{\rm b}, 0) + F_1'(n_{\rm b}, 0)
\over 2 N_{0 {\rm sym}}} \,,
\label{sym11} \\
f_1^{\rm np}(n_{\rm b}, 0) &=&
{F_1(n_{\rm b}, 0) - F_1'(n_{\rm b}, 0)
\over 2 N_{0 {\rm sym}}} \,.
\label{sym22}
\end{eqnarray}
Here $N_{0 {\rm sym}} \equiv N_{0 {\rm n}}(n_{\rm b},0)
=N_{0 {\rm p}}(n_{\rm b},0)$.
The plots of $F_1$ and $F_1'$
versus the wave number
$\kF_{\rm sym}=(3 \pi^2 n_{\rm b}/2)^{1/3}$
are given in Figures 14 and 17
of Ref. \cite{Jetal82}
for the Reid v6 and BJ v6 interactions, respectively.
We have fitted these functions by
simple analytical
formulas which are given in Appendix B.
Eqs. (\ref{f1nn}) -- (\ref{f1np}) yield
\begin{equation}
a(n_{\rm b}) = f_1^{\rm nn}(n_{\rm b},0)=f_1^{\rm pp}(n_{\rm b},0)
\,, \quad \quad c(n_{\rm b}) = f_1^{\rm np}(n_{\rm b},0) \,.
\label{ac}
\end{equation}
Now, considering the pure neutron matter,
we obtain
\begin{equation}
f_1^{\rm nn}(n_{\rm b}, 1) = {F_1(n_{\rm b}, 1)
\over N_{0 {\rm n}}(n_{\rm b},1)} \, .
\label{pure}
\end{equation}
The function $F_1(n_{\rm b}, 1) \equiv F_1(\kF_{\rm pure})$,
with $\kF_{\rm pure}=(3 \pi^2 n_{\rm b})^{1/3} $,
is plotted in Figures 20 and 22 of Ref. \cite{Jetal82}
for the Reid v6 and BJ v6 interactions, respectively.
The fitting formulas for $F_1(\kF_{\rm pure})$
are given in Appendix B.
From Eq. (\ref{f1nn}) we have
\begin{figure}[t]
\setlength{\unitlength}{1mm}
\leavevmode
\hskip  0mm
\includegraphics[width=120mm,bb=18  145  570  690,clip]{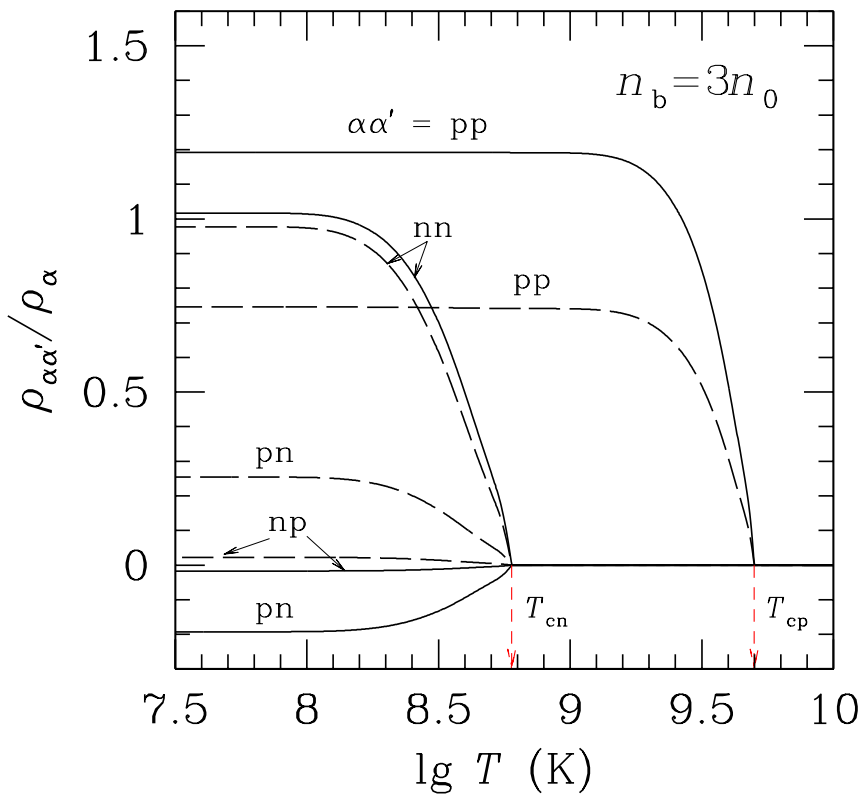}
\vspace{-1.0cm}
\caption
{The quantities $\rho_{\alpha \alpha'}/\rho_{\alpha}$
($\alpha, \alpha' = {\rm n}$ or ${\rm p}$)
versus temperature $T$ for two interactions:
BJ v6 (solid lines) and Reid v6 (long dashes)
at the baryon number density $n_{\rm b} = 3 n_{0}$.
The proton ($T_{c {\rm p}}=5 \times 10^9$~K)
and the neutron ($T_{c {\rm n}}=6 \times 10^8$~K)
critical temperatures are marked by the
vertical dashed arrows.}
\vspace{5mm}
\label{1}
\end{figure}
%
\begin{equation}
b(n_{\rm b}) = f_1^{\rm nn}(n_{\rm b},1) - a(n_{\rm b}) \,.
\label{b}
\end{equation}
Thus, assuming that $f_1^{\alpha \alpha'}$ is linear 
in $\delta$, one can calculate the Landau parameters
of the asymmetric nuclear matter and find
the entrainment matrix.

Our results are illustrated in Fig. 1.
We show the temperature dependence of
$\rho_{\alpha \alpha'}/\rho_{\alpha}$
($\alpha, \alpha' = {\rm n}$ or ${\rm p}$)
for the BJ v6 (solid lines) 
and Reid v6 (long dashes) interactions.
The number density of baryons is taken to be
$n_{\rm b} = 3 n_{0}$, where $n_0=0.16$ fm$^{-3}$ 
is the number density of nuclear matter at saturation.
For the chosen $n_{\rm b}$,
the equation of state of
Heiselberg and Hjorth-Jensen \cite{HH99}
yields $\delta = 0.837$.
The nucleon critical temperatures are taken to be
$T_{c {\rm n}} = 6 \times 10^8$ K,
$T_{c {\rm p}}= 5 \times 10^9$ K.
This information is sufficient for computing
the entrainment matrix at any $T$.
However, it should be stressed,
that our approach is not self-consistent.
Strictly speaking, one needs to calculate the
equation of state, the Landau parameters,
and the nucleon critical temperatures
using {\it one model of strong interactions}.

At $T \geq T_{c{\rm p}}$ in Fig. 1
the matter is non-superfluid and all the
matrix elements $\rho_{\alpha \alpha'} = 0$.
At $T_{c {\rm n}} \leq T \leq T_{c {\rm p}}$
protons become superfluid and hence $\rho_{\rm pp} \neq 0$.
Finally, at $T < T_{c {\rm n}}$ both 
protons and neutrons are superfluid
and all the matrix elements are non-zero.
With decreasing temperature,
the elements $\rho_{\alpha \alpha'}$ rapidly
approach their asymptotes (\ref{Borumand1}) and (\ref{Borumand2}).
Notice, that $\rho_{\rm pp}/\rho_{\rm p}$
depends essentially on the model of strong interactions employed
(the difference between the solid and dashed curves
marked by ${\rm pp}$ is large),
for $\rho_{\rm pn}/\rho_{\rm p}$ 
and $\rho_{\rm np}/\rho_{\rm n}$
the dependence is even stronger:
even the signs of these quantities
differs for the BJ v6 and Reid v6 interactions
(for the Reid v6 interaction $\rho_{\rm pn}=\rho_{\rm np} <0$
because $f_1^{\rm np}<0$; see Eq. (\ref{Borumand2})).
The quantity $\rho_{\rm pp}/\rho_{\rm p}$
at $T \rightarrow 0$
is equal to $\rho_{\rm pp}/\rho_{\rm p} \approx 1.19$
for the BJ v6 interaction
and to $\rho_{\rm pp}/\rho_{\rm p} \approx 0.75$  for
the Reid v6 interaction.
Thus, the value of $\rho_{\rm pp}/\rho_{\rm p}$
is approximately
half of that obtained
in the estimate of Borumand et al. \cite{Betal96},
$\rho_{\rm pp}/\rho_{\rm p} \approx 2$.

\section{Conclusions}
We have derived the entrainment matrix
$\rho_{\alpha \alpha'}$ at non-zero temperatures.
The calculation is performed in the frame
of the Landau Fermi-liquid theory generalized
by Larkin and Migdal \cite{LM63} and Leggett \cite{L65}
to the case of superfluid matter.
The expressions for $\rho_{\alpha \alpha'}$
reproduce two limiting cases 
studied in the literature:
the case of zero temperature and the case in which
the interaction between neutrons and protons is absent.

The results are presented in the form convenient for their
practical use.
In particular, for calculating the entrainment matrix
one needs the Landau parameters of symmetric nuclear
and pure neutron matter.
These parameters were
taken from the paper by Jackson et al. \cite{Jetal82}
for two model potentials of
nucleon-nucleon interaction (BJ v6 and Reid v6)
and approximated by simple analytical formulas.
While results for $\rho_{nn}$ are quite similar for both
nucleon-nucleon potentials, the values of $\rho_{pp}$ differ
by some thirty percent, while much smaller non-diagonal matrix
elements,   $\rho_{np}$  and $\rho_{pn}$ differ in sign.
Clearly, results for the non-diagonal entrainment matrix
elements are very sensitive to the nucleon-nucleon
interaction. Moreover, as we used old values of the Landau
parameters, our results should be updated as soon as new, more
realistic values of the $l=1$ parameters become available,
obtained, e.g., via modern renormalization  group approach
(see, e.g., Ref. \cite{Setal03})
applied to pure neutron matter, symmetric nuclear matter, and
maybe also to the most astrophysically interesting case of
asymmetric nuclear matter.

The generalization of the entrainment matrix to the case of
three or more interacting baryon species is straightforward.
It just implies an increased corresponding number of
baryon indices in all formulas and an associated
increase of the dimension of matrices 
${\pmb \rho}$ and ${\pmb \gamma}$.

The {\it temperature dependence}
of the matrix $\rho_{\alpha \alpha'}$,
fully described by
the universal function $\Phi_{\alpha}(T)$,
is known rather reliably. Unfortunately,
this cannot be said about the values of the
Landau parameters.
Clearly, new calculations of the 
Landau parameters of asymmetric nuclear matter
would be highly desirable.

The matrix $\rho_{\alpha \alpha'}$ at non-zero
temperatures is needed to study 
the kinetics of the neutron star matter
as well as to investigate 
the dynamical evolution of neutron stars,
especially, their pulsations.
As we have shown, the entrainment matrix
varies with temperature. 
Because the matrix $\rho_{\alpha \alpha'}$ enters
hydrodynamic equations
which determine the neutron star pulsations,
the frequencies of (superfluid) pulsation modes
should vary with $T$ and hence
(due to a star cooling or heating), with time.
This gives a potentially powerful method to probe very
subtle properties of superdense matter by measuring
the dependence of pulsation frequencies on time.
We intend to consider the related problems
in a separate publication.

\begin{ack}
The authors are grateful to D.G. Yakovlev for discussion.
One of the authors (M. Gusakov)
also acknowledges the excellent
working condition at the
N. Copernicus Astronomical Center in Warsaw,
where this study was completed.

This research was supported
by RFBR (grants 03-07-90200 and 05-02-16245),
the Russian Leading Science School (grant 1115.2003.2),
INTAS YSF (grant 03-55-2397), 
the Russian Science Support Foundation,
and by the Polish MNiI Grant No. 1 P03D 008 27.
\end{ack}

\appendix
\section{The analytical approximation of the function~$\Phi_{\alpha}$}
\label{appendix1}
Eq. (\ref{N0}) for $\Phi_{\alpha}$
can be written as:
\begin{equation}
\Phi_{\alpha} = 2 \, \int_0^{\infty} \, \dd x \,
{ \exp \left(\sqrt{x^2 + v^2} \right)    \over \left[
\exp\left(\sqrt{x^2 + v^2} \, \right) + 1 \right]^2} \,.
\label{intPhi}
\end{equation}
Here $v = \Delta(T)/T$; $\Delta(T)$ is the
gap which depends on the type of pairing.
For singlet-state pairing of nucleon species $\alpha$
the function $v$ can be taken from Levenfish and Yakovlev \cite{LY94}:
\begin{equation}
v = { \Delta^{(\alpha)}(T) \over T} =
\sqrt{1- \tau} \, \left( 1.456 - {0.157 \over \sqrt{\tau} } + {1.764 \over \tau}
\right)
\, , \quad \quad \tau = {T \over T_{c \alpha} } \,.
\label{singlet}
\end{equation}
In the case of triplet-state neutron pairing the effective
gap should be introduced which is defined by Eq. (\ref{effGAP}).
The function $v$ is now given by
(see Ref. \cite{YL95}):
\begin{equation}
v = { \Delta^{(\rm n)}_{\rm eff}(T) \over T} =
\sqrt{1- \tau} \, \left( 0.7893 + {1.188 \over \tau} \right)
\, , \quad \quad \tau = {T \over T_{c {\rm n}} } \,.
\label{triplet}
\end{equation}
Gnedin and Yakovlev \cite{GY95}
calculated the function $\Phi_{\alpha}(v)$
in a wide range of $v$ 
(for the problem of thermal conductivity).
These authors fitted $\Phi_{\alpha}(v)$ by a simple
analytical formula which is correct at any $v$
and satisfies the asymptotes
$\Phi_{\alpha} ~=~\sqrt{2 \pi v} \, \, {\rm e}^{-v}$
at $v \rightarrow + \infty$:
\begin{equation}
\Phi_{\alpha} = \left[0.9443 + \sqrt{(0.0557)^2 + (0.1886 v)^2} \, \right]^{1/2}
\,\, \exp \left( 1.753 - \sqrt{(1.753)^2 + v^2} \right) \, .
\label{Fit1} \\
\end{equation}
Calculation and fit errors do not exceed $2.6 \%$.

\section{Fits to $F_1$ and $F_1'$}
\label{appendix2}

\subsection{Symmetric nuclear matter}
In the symmetric nuclear matter $\delta=0$.
The plots of functions $F_1$ and $F_1'$ versus the wave number
$\kF_{\rm sym}=(3 \pi^2 n_{\rm b}/2)^{1/3}$ are given
by Jackson et al. \cite{Jetal82} for the model interactions BJ~v6 and Reid~v6
(see Section 4.2 for details).
We have approximated these functions by simple analytical
formulas in the interval
$1.2 \,{\rm fm}^{-1} \leq \kF_{\rm sym} \leq 2.0 \,{\rm fm}^{-1}$.
For the BJ v6 interaction we obtain
\begin{eqnarray}
F_1(n_{\rm b}, 0) &=& -0.6854 + 0.6724 \, \kF_{\rm sym} - 0.5180 \, (\kF_{\rm
sym})^2 \,,
\label{F1_BJ} \\
F_1'(n_{\rm b}, 0) &=& 1.723 - 1.520 \, \kF_{\rm sym} + 0.03498 \, (\kF_{\rm
sym})^2 \,,
\label{F1'_BJ}
\end{eqnarray}
while for the Reid v6 interaction
\begin{eqnarray}
F_1(n_{\rm b}, 0) &=& 1.034 -1.866 \, \kF_{\rm sym} + 0.5455 \, (\kF_{\rm
sym})^2 \,,
\label{F1_R} \\
F'_1(n_{\rm b}, 0) &=& 0.6973 + 0.1403 \, \kF_{\rm sym} - 0.5303 \, (\kF_{\rm
sym})^2 \,.
\label{F1'_R}
\end{eqnarray}
%
\subsection{Pure neutron matter}
In the pure neutron matter $\delta=1$.
The plots of the function $F_1$ versus the wave number
of the pure neutron matter
$\kF_{\rm pure}=(3 \pi^2 n_{\rm b})^{1/3}$ are given by
Jackson et al. \cite{Jetal82} (see also Section 4.2).

The fit for the BJ v6 interaction has the form
\begin{equation}
F_1(n_{\rm b}, 1)
= 0.1473 + 0.7372 \, \kF_{\rm pure} -1.0414 \, (\kF_{\rm pure})^2 + 0.1958 \,
(\kF_{\rm pure})^3 \,,
\label{F1_BJpure}
\end{equation}
while for the Reid v6 we obtain
\begin{equation}
F_1(n_{\rm b}, 1) = -0.2729 + 1.5545 \, \kF_{\rm pure} -1.3225 \,
(\kF_{\rm pure})^2  + 0.2393 \, (\kF_{\rm pure})^3
\,.
\label{F1_Rpure}
\end{equation}
The fitting formulas (\ref{F1_BJpure}) and (\ref{F1_Rpure})
correctly describe the function $F_1(n_{\rm b}, 1)$
in the interval $0.75 \, {\rm fm}^{-1} \leq \kF_{\rm pure}
\leq 3.0 \, {\rm fm}^{-1}$.


\end{document}